\documentclass[final,3p,times,twocolumn]{elsarticle}
\usepackage{graphics}
\usepackage{graphicx}
\usepackage{epsfig}

\usepackage{amssymb}
\usepackage{amsthm}

\usepackage{multirow}
\usepackage{multicol}

\usepackage{hyperref}

\usepackage[normalem]{ulem}
\usepackage{color}
\definecolor{blue}{rgb}{0,0,1}
\definecolor{red}{rgb}{1,0,0}

\usepackage{array}
\newcolumntype{C}[1]{>{\centering\let\newline\\\arraybackslash\hspace{0pt}}m{#1}}

\usepackage{lineno}

\biboptions{comma, square}

\journal{Nuclear Instruments and Methods in Physics Research A}

\begin{document}

\begin{frontmatter}


%% \tnotetext[label1]{Footnote to title.}

\title{Low Momentum Particle Detector for the NA61 Experiment at CERN}
%% \tnoteref{titlefootnote}}
%% \tnotetext[titlefootnote]{NA61/SHINE is a fixed target experiment at the CERN SPS [\texttt{http://na61.web.cern.ch}].}


\author[rmki]{Krisztina M\'arton}
\ead{marton.krisztina@wigner.mta.hu}

\author[rmki,elte]{G\'abor Kiss}
\ead{kiss.gabor@wigner.mta.hu}

\author[rmki]{Andr\'as L\'aszl\'o}
\ead{Andras.Laszlo@cern.ch}

\author[rmki]{Dezs\H o Varga}
\ead{Dezso.Varga@cern.ch}

\address[rmki]{Institute for Particle and Nuclear Physics, MTA Wigner Research Centre for
Physics, Budapest, Hungary}
%% \address[cern]{CERN, CH-1211 Geneva 23, Switzerland}
\address[elte]{E\"otv\"os University, Budapest, Hungary}

\begin{abstract}

The NA61 Experiment at CERN SPS is a large acceptance hadron spectrometer, aimed to studying of hadron-hadron, hadron-nucleus, and nucleus-nucleus interactions in a fixed target environment. The present paper discusses the construction and performance of the Low Momentum Particle Detector (LMPD), a small time projection chamber unit which has been added to the NA61 setup in 2012. The LMPD considerably extends the detector acceptance towards the backward region, surrounding the target in  hadron-nucleus interactions. The LMPD features simultaneous range and ionization measurements, which allows for particle identification and momentum measurement in the 0.1 -- 0.25 GeV/$c$ momentum range for protons. The possibility of Z=1 particle identification in this range is directly demonstrated.

\end{abstract}

\begin{keyword}
CERN NA61 \sep TPC \sep centrality measurement \sep gray proton

\end{keyword}

\end{frontmatter}



\section{Introduction}
\label{introduction}

Over the last four decades of experimental study of hadronic
interactions, a large amount of information has been gathered on
production of ``slow'' particles, which are slow in the sense that
in a fixed target environment their rapidity in the target frame is less
than unity. The term limiting fragmentation \cite{limfrag} has
actually been formulated for this region, and scaling properties have
been studied for large variety of reactions.

Considering particle production from a target nucleus in a fixed target hadron-nucleus
(h+A) or nucleus-nucleus (A+A) interactions with beam energies in the order 
of a few GeV, a sizeable low-energy
component emerges due to the de-excitation of the nucleus: nucleons or
smaller nuclei are produced with kinetic energy of the order of the
nuclear binding energy. This component is generally referred to as
``black'', a name which originates from early emulsion studies \cite{Heckman}. There
is an other component, which is strongly connected to the
fragmentation of nucleons and is attributed to intra-nuclear
cascading. These ``gray'' particles, mostly nucleons but also pions
and light nuclei, carry kinetic energy of 30 -- 400 MeV, considerably
higher than the nuclear binding energy (for a complete review, see \cite{Fredriksson}). 
Finally, there is an additional 
component of the slow particles which
resembles that observed in h+p or h+n collisions \cite{na49pc_disc}, 
such as particles in the diffractive peak.

The LMPD (Low Momentum Particle Detector), an integral part of the NA61 
Experiment \cite{na61proposal}, 
aims at differentiating centrality in p+A and A+A interactions
recorded by the CERN experiment NA61, and to clarify quantitatively
the details of the relation between event centrality and slow particle
production. NA61 has a key advantage having high acceptance, allowing
identified produced particles (including strangeness content, central baryons
and antibaryons) and slow particles (by LMPD) to be measured by the
same apparatus \cite{na61facility}.

\subsection{Centrality control in h+A}
\label{centralitycontrol}

The production of low momentum particles in high energy hadron-nucleus 
collisions were studied by many experiments over the last few decades \cite{Sikler}. 
A key observation was that the number of the slow nucleons, 
especially in the ``black'' and ``gray'' regions emerging from the break-up of the nucleus, 
gives information about the 
centrality (the impact parameter) of the h+A collision. 

The h+A collisions were studied at various energies, with different types 
of projectiles and targets.
It was found that the angular distributions of the low momentum (``gray'' 
and ``black'') protons are to first order independent of the energy and 
of the type of the incoming projectile, but they show significant dependence on the mass of 
the target nucleus: they are stronger forward-peaked for lighter targets. The angular distributions 
for ``gray'' protons are forward-peaked, while for the ``black'' ones show only little 
asymmetry \cite{Heckman, WA35, E667}.

Regarding p+C interactions, a comprehensive data survey with critical review of compatibility 
between various measurements has been recently published \cite{NA49backwards}, 
incorporating relevant new measurements by the NA49 Experiment in p+C interactions \cite{na49pc_pr}.
This completes earlier discussion of the p+C collision system \cite{na49pc_disc}, clarifying the
momentum regions populated by the different production mechanisms.

The yield of these slow protons in the h+A interaction is found to have two sources: the nucleon-nucleon 
encounters and the processes involving nuclear matter. This latter group of processes have also important 
role in the production of deuterons, tritons and other light nuclei \cite{Piroue}.

It was suggested already in 1976 \cite{Gutbrod} that by measuring the 
large composite fragments, one can select the central collisions, and subsequently 
it was proposed \cite{Yaeger} that the number of heavily ionizing particles 
$N\sb{h}$ measures the number of struck nucleons inside the target nucleus. 

The energy independence of the distribution of these heavy particles 
supported the hypothesis that $N\sb{h}$ measures the impact parameter 
(the centrality) of the h+A collision and it is correlated to the number 
of nucleon-nucleon collisions in the nucleus \cite{E668}. With increasing 
centrality the number of ``gray'' nucleons increases almost proportionally, 
whereas the number of ``black'' particles saturates for 
central collisions \cite{EMU07}.

One of the aims of the LMPD is to understand the transition from ``black''
to ``gray'' energy ranges, and to quantify how these regions are related to centrality. Our
choice of technology matches this region, with best performance
(proton identification and momentum measurement) in the 15 -- 30 MeV
kinetic range, well covering the transiton from ``black'' to ``gray''. ``Gray''
protons up to 500 MeV/$c$ momentum (120 MeV kinetic energy) can be tagged by
their high ionization deposit.

\subsection{NA61 environment}
\label{NA61enviroment}

\begin{figure}[!ht]
\begin{center}
\includegraphics[angle=0, width=7.5cm]{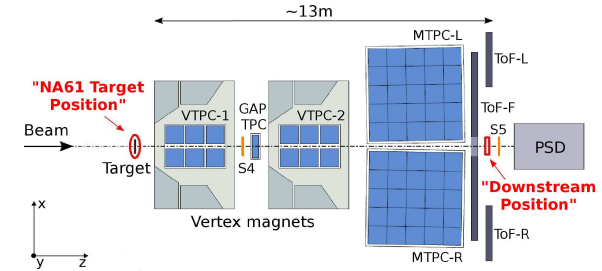}
\caption{Outline of the NA61/SHINE Experiment. LMPD data taking positions are indicated.}
\label{fig:na61setup}
\end{center}
\end{figure}

During normal physics data taking in 2012, the LMPD was an integrated part of
the NA61 detector system. The position of the LMPD for these periods
is indicated in Figure~\ref{fig:na61setup}, surrounding the target. Also, data 
has been taken for a considerable time in 2011 in
a downstream position, to exploit the available beam time for
configurations incompatible with the LMPD (use of hydrogen target). In this
case the LMPD was independent from the rest of the NA61 system, with a
stand-alone trigger, data acquisition, and target setup.

\begin{table}[!h]
\caption{Collected data at 158 GeV/c in ``downstream'' position (2011, standalone) and in ``NA61 target'' position (2012, full NA61).}
\begin{center}
\begin{tabular}{|C{1cm}|C{3cm}|C{2.5cm}|}
\hline
\textsl{Year} & \textsl{Target} & \textsl{Collected events} \\ \cline{1-3}
\multirow{5}{*}{2011} & Pb, 0.5mm & $2 \, 442$ k \\  
& Pb (rotated), 0.5mm & $617$ k \\  
& C, 2mm & $547$ k \\ 
& Al, 1mm & $622$ k \\ 
& Target out & $264$ k \\ \cline{1-3}
\multirow{5}{*}{2012} & $\sp{83m}$Kr (calibration) & $1 \, 593$ k \\ \cline{2-3}
& Pb, 0.5mm & $2 \, 140$ k \\ 
& Target out & $274$ k \\ \cline{2-3}
& Pb, 1mm & $9 \, 206$ k \\ 
& Target out & $927$ k \\ \cline{1-3}
\end{tabular}
\end{center}
\label{adatok}
\end{table}

This downstream position, behind the MTPCs (see also
Figure~\ref{fig:na61setup}) allowed a flexible change of operational
conditions, therefore most of the technical studies were performed
here.

The integration of the LMPD unit into the NA61 environment was
largely simplified by the fact that the LMPD uses the same front-end
electronics as the existing NA61 TPCs. The detector has been included
in NA61 data acquisition system and the online monitoring system as well, in a
fashion compatible with all the other TPC units.

\section{Detector construction}
\label{detconst}

\subsection{Principle of operation}
\label{principle}

 The detector exploits the simultaneous measurements of ionization
 (dE/dx) and range, which, due to the different mass, makes a
 differentiation between particle types. The range measurement is
 rough, typically with a precision of a factor of two; this is however
 sufficient to specify a narrow momentum bin, since the momentum dependence of range is very 
 steep (approximately proportional to the fourth power of the momentum). The
 ionization ratio at a given range for any two types of particles is
 approximately proportional to the square root of the mass ratio: this
 implies that also the dE/dx measurement need not to be very precise
 for clear identification (between pions and protons the ionization ratio is around 2.6 for a given range). 
 For low momentum particles of interest the
 ionization is high (about 5-20 times the minimum), allowing sampling in a
 gas gap of a few cm. 
 In addition, the per event multiplicity of these particles is rather low, 
 up to a few tens with an approximately spherical distribution.
 These considerations led to a rather compact
 detector outline, where position sensitive detection layers are
 interspersed with absorber layers. The thin detector walls imply operation 
 at atmospheric pressure, specifically, about 0.2 - 0.5 mbar above ambient 
 pressure.

 The actual design was guided by a simulation based on the Photon
 Absorption Ionization (PAI) \cite{pai} model. The comparison between
 the simulation and the measurements are discussed in detail in Section
 ~\ref{comparison}.

\subsection{Detector outline}
\label{detoutline}

\begin{figure}[!ht]
\begin{center}
\includegraphics[angle=0, width=6.0cm]{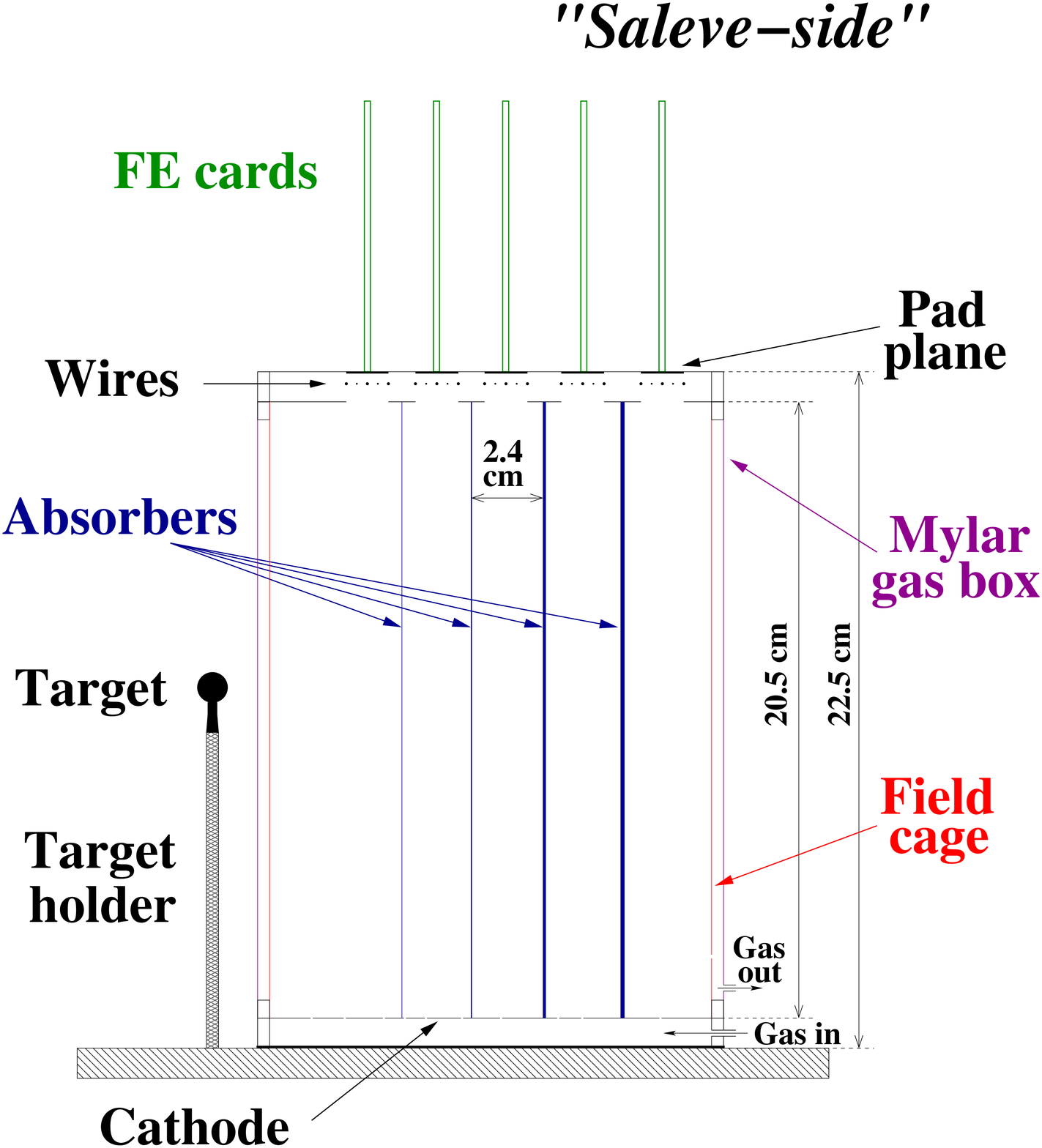}
\caption{Detector outline (one half) from the beam direction.}
\label{fig:principle_outline}
\end{center}
\end{figure}

The Low Momentum Particle Detector is a small time projection chamber 
with absorber layers in the gas volume. The detector outline following 
the principles discussed in Section~\ref{principle} is shown in 
Figure~\ref{fig:principle_outline}. The absorber layers define intervals in 
the range of the particles and they also act as an inner field cage. The 
vertical electric field in the LMPD guides the produced ionization electrons 
drifting towards the top of the detector, where they are read out by a 
multi wire proportional chamber.

\begin{figure}[!ht]
\begin{center}
\includegraphics[angle=0, width=7.0cm]{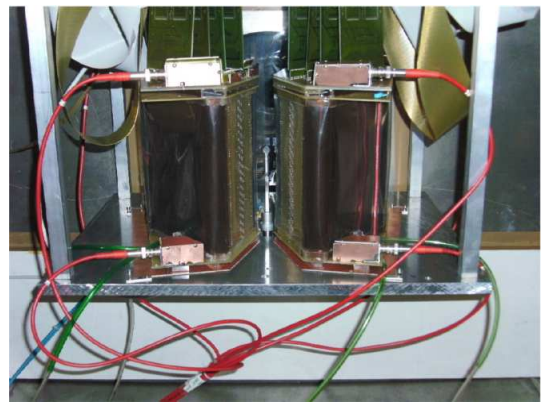}
\caption{Complete system (LMPD+target) in ``NA61 target'' position.}
\label{fig:lmpd_photo}
\end{center}
\end{figure}

\begin{figure}[!ht]
\begin{center}
\includegraphics[angle=0, width=7.0cm]{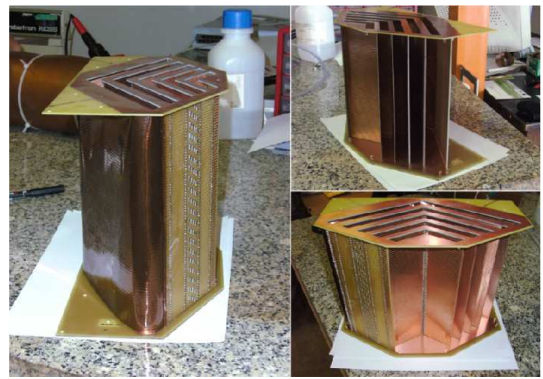}
\caption{LMPD: absorber layers and field cage of ``Jura-side''.}
\label{fig:LMPD1}
\end{center}
\end{figure}

LMPD has two independent parts, the ``Jura''- and the ``Saleve''-sides, 
see Figure~\ref{fig:lmpd_photo}. Figure~\ref{fig:LMPD1} shows the absorber 
layers and field cage of ``Jura-side''. The absorbers are glass-epoxy (G10) 
sheets with 2~mm wide horizontal Cu strips. The outer field cage is a
60~$\mu$m kapton foil printed with 5~$\mu$m Cu strips. The readout MWPC 
has approximately radial pad structure. There are 10 pad rows, 
the absorbers are after every second pad row, defining 5 detection layers. (More details about the
readout chamber are in Section~\ref{readoutmwpc}.)

In 2010 a prototype of LMPD was also built. This ``2010 Proto'' unit 
was found to be useful as a multiplicity monitor in the downstream setup, 
in combination with the final detector.

\subsection{Gas system}
\label{gassystem}

LMPD has double walls, similarly to the other TPC chambers of
the NA61 detector. The inner wall is a 60~$\mu$m kapton layer, the
outer one is a 40~$\mu$m thick mylar foil. Mixture of 85\% Ar + 15\% CO$\sb{2}$
has been used as filling gas. 

The gas enters from the bottom part of the chamber through 30 holes 
drilled in the cathode plane, with
1 mm diameter each, in order to evenly distribute the fresh gas between the
absorber layers. The used gas is then guided to fill the layer between the 
kapton and the mylar foil before being vented from the chamber. 
This solution allows one to exploit the quality improvement achieved by the
double wall structure, without the need of an additional gas circulation
path.

\subsection{Absorber structure}
\label{absorber}

The absorber structure of LMPD is shown in Figure~\ref{fig:LMPD1}. 
There are 4 absorber layers in both Jura- and Saleve-sides, 
placed after every second pad row. The absorbers are made from
glass-epoxy. Their thicknesses are 0.5, 1.0, 2.0, and 2.5~mm,
however the effective thickness depends on the angle of incidence. 
The detector wall acts as an absorber, and hence defines the minimum 
detectable particle energy.

As a general overview, Table~\ref{tab:absorbers} gives the basic
properties of the absorbers, as well as the momentum cutoff for
protons which are able to pass through the given absorber
layer. These latter quantities, especially the ionization, have a
complicated dependence on the particle and detector geometry as well
as the energy distribution, therefore will be subject of a detailed
analysis. The key message of the present paper is to demonstrate the
possibility of a clean measurement for these approximate kinematic
ranges.

\renewcommand{\arraystretch}{1.25}
\begin{table*}[!hbt]
\caption{Absorber thicknesses, approximate momentum ranges and ionization (in 1.2 cm Ar) for perpendicular incidence.}
\begin{center}
\begin{tabular}{|C{2.1cm}|C{2.1cm}|C{2.1cm}|C{2.1cm}|C{2.1cm}|C{2.1cm}|}
\hline
\textsl{Absorber number} & \textsl{Thickness} (mm) & \textsl{Thickness} ($\mathrm{g/cm^2}$) & \textsl{Cumulative thickness} ($\mathrm{g/cm^2}$) &\textsl{Momentum cutoff} (MeV) & \textsl{Most probable ionization} (keV)\\ \hline
Detector wall & 0.1 & 0.018 & 0.018 & 110 & -- \\ \hline
1             & 0.5 & 0.106 & 0.124 & 127 & 100 \\ \hline
2             & 1.0 & 0.213 & 0.337 & 171 & 63 \\ \hline
3             & 2.0 & 0.425 & 0.762 & 216 & 38 \\ \hline
4             & 2.5 & 0.532 & 1.294 & 250 & 26 \\ \hline
\end{tabular}
\end{center}
\label{tab:absorbers}
\end{table*}

\subsection{Readout MWPC}
\label{readoutmwpc}

\begin{figure}[!ht]
\begin{center}
\includegraphics[angle=0, width=6.0cm]{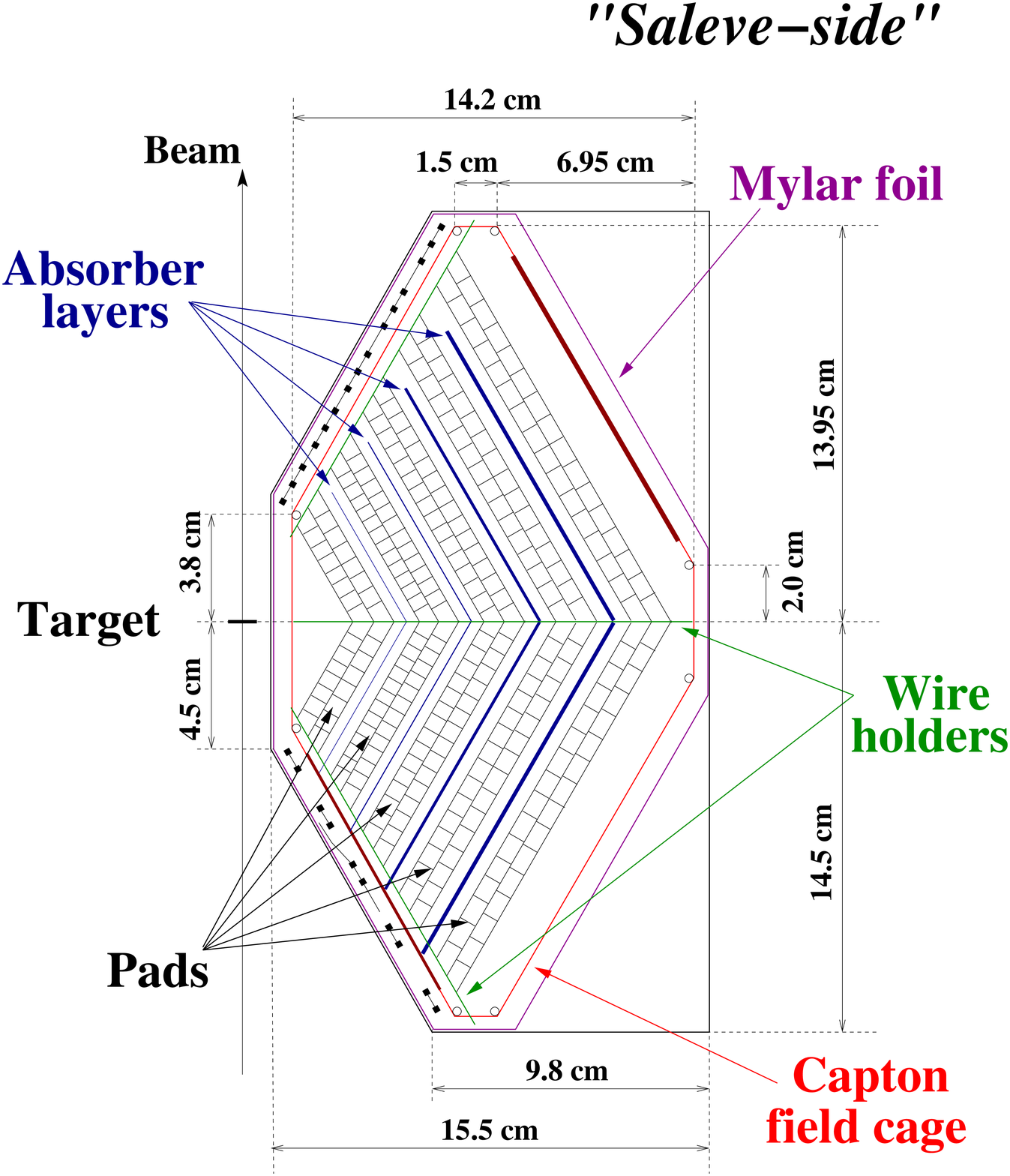}
\caption{Pad-structure of the ``Saleve-side'' detector.}
\label{fig:LMPD_wedge}
\end{center}
\end{figure}

The readout chamber is a MWPC which is placed on the top of the
chamber, with segmented cathode (pads) on ground potential.
The close to radial pad-structure is shown in Figure~\ref{fig:LMPD_wedge}. 
The pads are organized in 10 rows perpendicularly to the typical track direction, the 
number and size of pads increase towards the outer pad rows.

In the readout chamber there are two kind of wires, the sense (anode) wires
with 21~$\mu$m thickness and the field wires (100~$\mu$m thick). 
The wires are to first order compatible with the radial structure of pads with the help of a
wire-holder in the middle of the pad plane, which bends the wires
on a short section and therefore reduces the overall dead zone. The
wire-holder in the middle divides the pad-structure to two symmetric parts
(``wedges'', see Figure~\ref{fig:LMPD_wedge}). These wedges are handled 
independently during the analysis.

The pads are 6~mm long, that is, the track segment between each
absorber pair is measured on a 12~mm segment. The distance between
sense wires are correspondingly 6~mm. The distance between the wire plane
and the pad plane is 4~mm.

The signal formation is based on the same principle as for the larger
TPC-s of the NA61 detector \cite{na61facility, na49}, with avalanche formation on the
sense wires, and capacitive signal coupling to the pads.

Since the LMPD detects highly ionizing slow particles, the optimal gas
multiplication gain is below the typical TPC gains designed for
minimum ionization. This implies that even gating grid is not
necessary, which would otherwise reduce ion backflow to the TPC
sensitive volume. The total current measured on the 
sense wire high voltage supply line was typically 20~nA at full 
beam rate, that is, around 0.07~nA/cm current density, which justified 
this approach.

\begin{figure}[!ht]
\begin{center}
\includegraphics[angle=0, width=7.5cm]{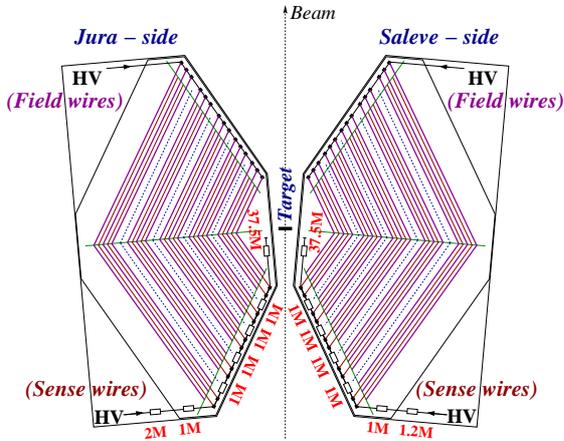}
\caption{Top view of the LMPD: mechanical and high voltage support for the wires, 
with the resistor chain on the sense wires indicated.}
\label{fig:reschain}
\end{center}
\end{figure}

The typical proton momentum is lower for those tracks which stop
early, and higher for those which run along all the detection
layers. In order to optimize the electronics dynamic range, a
gradually increasing gas multiplication gain has been applied towards
outer pad rows to follow the decreasing ionization due to increasing
momentum. The practical realization relied on a resistor chain 
(see Figure~\ref{fig:reschain}), with a constant voltage drop between 
each absorber layers for the sense wires.

\begin{table}[!h]
\caption{High voltage settings during the physics run.}
\begin{center}
\begin{tabular}{|C{2.4cm}|C{1.7cm}|C{1.7cm}|}
\hline
 & \textsl{Saleve-side} & \textsl{Jura-side} \\ \hline
Cathode HV & -4000 V & -4000 V \\ \hline
Field Wire HV & -400 V & -400 V \\ \hline
Sense Wire HV & 1150 V & 1150 V \\ \hline
Pad row 1,2 - SW & 987 V & 969 V \\ \hline 
Pad row 3,4 - SW & 1013 V & 995 V \\ \hline
Pad row 5,6 - SW & 1040 V & 1021 V \\ \hline
Pad row 7,8 - SW & 1066 V & 1047 V \\ \hline
Pad row 9,10 - SW & 1092 V & 1073 V \\ \hline
\end{tabular}
\end{center}
\label{HVsettings}
\end{table}

\subsection{Simulation of electron drift}
\label{simulation}

In order to find the appropriate voltage settings, electrostatic simulations 
have been performed with Garfield~\cite{Garfield}. As the geometrical 
properties of LMPD require thin wires and relatively large planes as well, 
the nearly exact Boundary Element Method 
solver \cite{neBEM} was used to calculate the electric field (Garfield is 
interfaced with the \emph{neBEM} program). 

\begin{figure}[!ht]
\begin{center}
\includegraphics[angle=0, width=7.5cm]{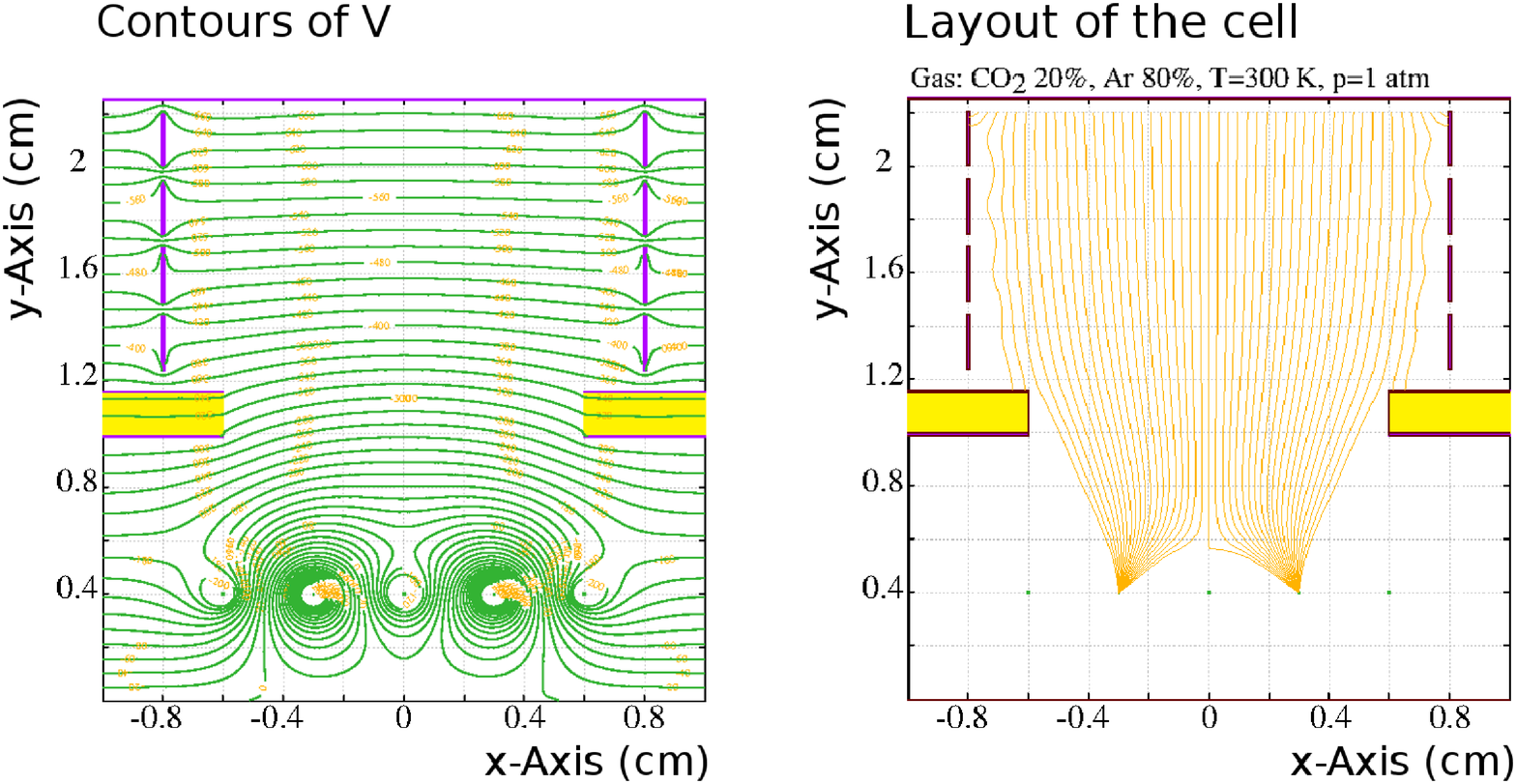}
\caption{Left panel: equipotential lines in the LMPD amplification cell (two pad rows out of ten). Right panel:drift lines of electrons.}
\label{fig:sim}
\end{center}
\end{figure}

Figure~\ref{fig:sim} shows the equipotential lines in the vicinity of the 
wire plane, as well as part of the field cage. The absorber walls are vertical (y coordinate), and the
wires are perpendicular to the plane of the Figure. On the right side of 
Figure~\ref{fig:sim} several calculated electron drift lines are shown, 
starting from $y=2.2$~cm position. This demonstrates that with 
these voltage settings the majority of electrons are collected 
by the anode wires. Note that further from the amplification cell, the
field structure is completely defined by the field cage, that is, the 
voltage settings have no effect on the collection efficiency.

The simulations confirmed the approach in which a single wire layer was
installed, simplifying the construction step. This implied however,
that the field wires are set on a considerable negative voltage, -400 V for
all field wire voltages. Such setting reduced the gain dependence on
cathode flatness \cite{ccc}, and thus improved gain uniformity.

%The drift lines have been calculated in Garfield with the Runge-Kutta-Fehlberg integration method. 

\subsection{Read-out and electronics}
\label{electronics}

Electronic signals from each of the individual cathode segments (pads) in the readout MWPC
are recorded by the same front-end (FE) cards as used for the NA61 tracking TPCs \cite{Kleinfelder,Bieser}. 
Each of these FE cards can store analog time trace of 32 TPC pads, 
with time sampling in 256 elements of 200~ns spacing, allowing total drift
time of 51.2 $\mu$s.
After sampling, the FE cards digitize the signals in a serial 
way using an on-card Wilkinson ADC. LMPD uses 18 such FE cards. 

The steering logic for the FE readout process is hosted on the 
readout mother boards (MB). The 9 bit pad charge ADCs from the FE cards are 
pedestal subtracted, truncated to 8 bit, noise suppressed and zero compressed 
by the MB before serializing them to an LVDS connection line towards 
a concentrator box (CB). These further serialize the data to a DDL optical 
connection line \cite{Rubin,Carena} towards the Central DAQ computer of the NA61 experiment. One 
MB can host up to 24 pieces of FE cards, thus only one is used 
for the LMPD (including all subunits).

The detected signal shapes, timing and the
noise performance was compatible with that experienced at the other
NA61 TPCs. The electronics control and maintenance (including regular
pedestal measurements, monitoring of power supply and data stream) was
integrated into the NA61 framework.

\begin{figure}[!ht]
\begin{center}
\includegraphics[angle=0, width=7.5cm]{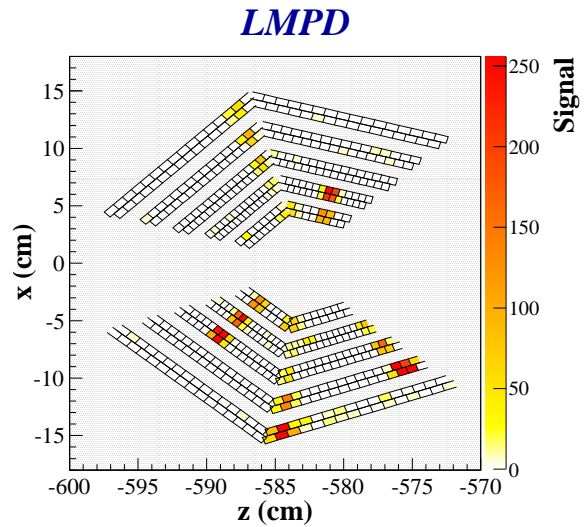}
\caption{Typical raw event of LMPD (top view).}
\label{fig:topview}
\end{center}
\end{figure}

Figure~\ref{fig:topview} shows event display of ADCs of a typical raw event, available for on-line
performance checks during the measurement.

\section{Target and trigger system}
\label{targettriggersystem}

\subsection{Trigger counters}
\label{triggercounters}

The detector by design operates with a target which is as thin and as narrow as possible. 
To reduce background, various
trigger counters were arranged in an optimum way.

\begin{figure}[!ht]
\begin{center}
\includegraphics[angle=0, width=7.5cm]{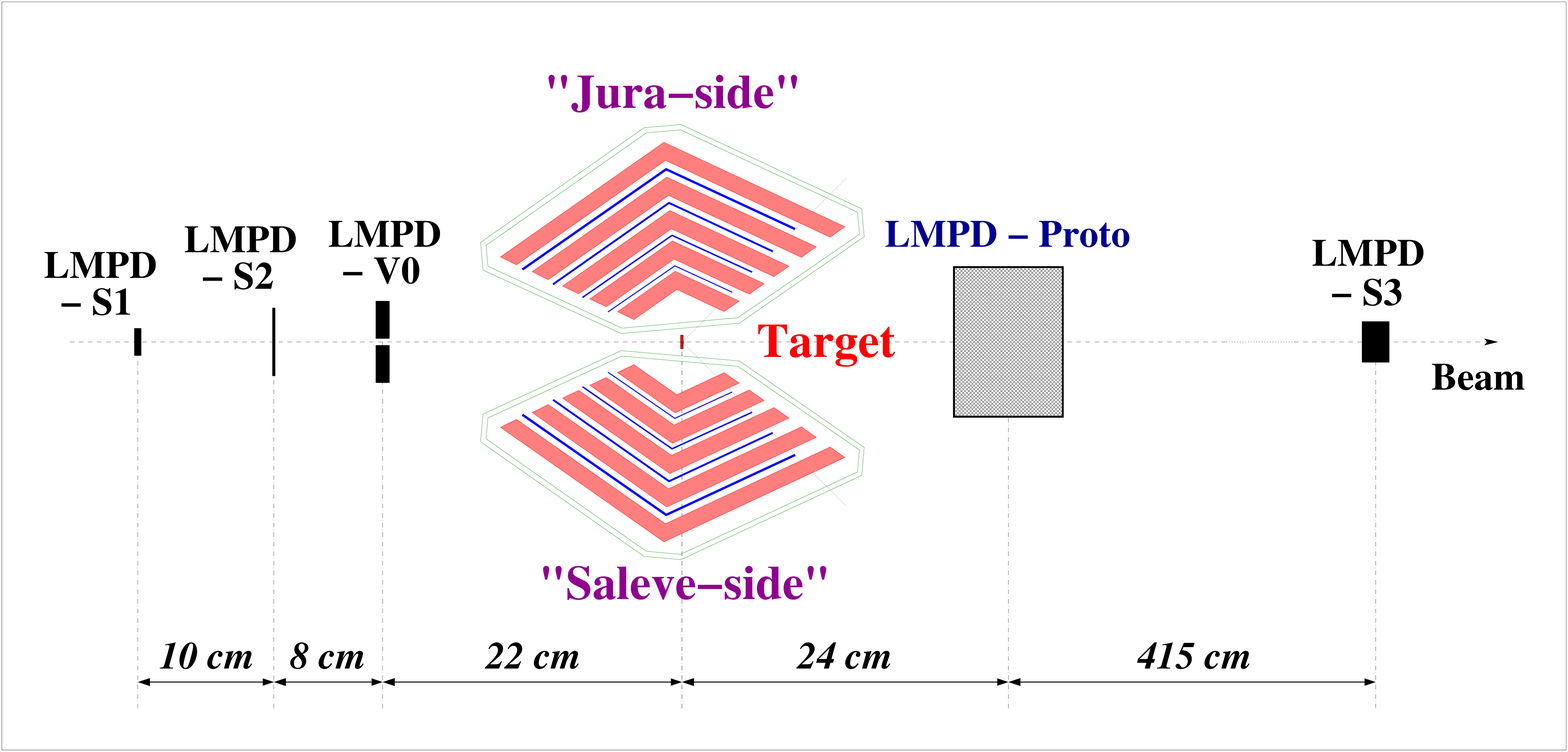}
\includegraphics[angle=0, width=7.5cm]{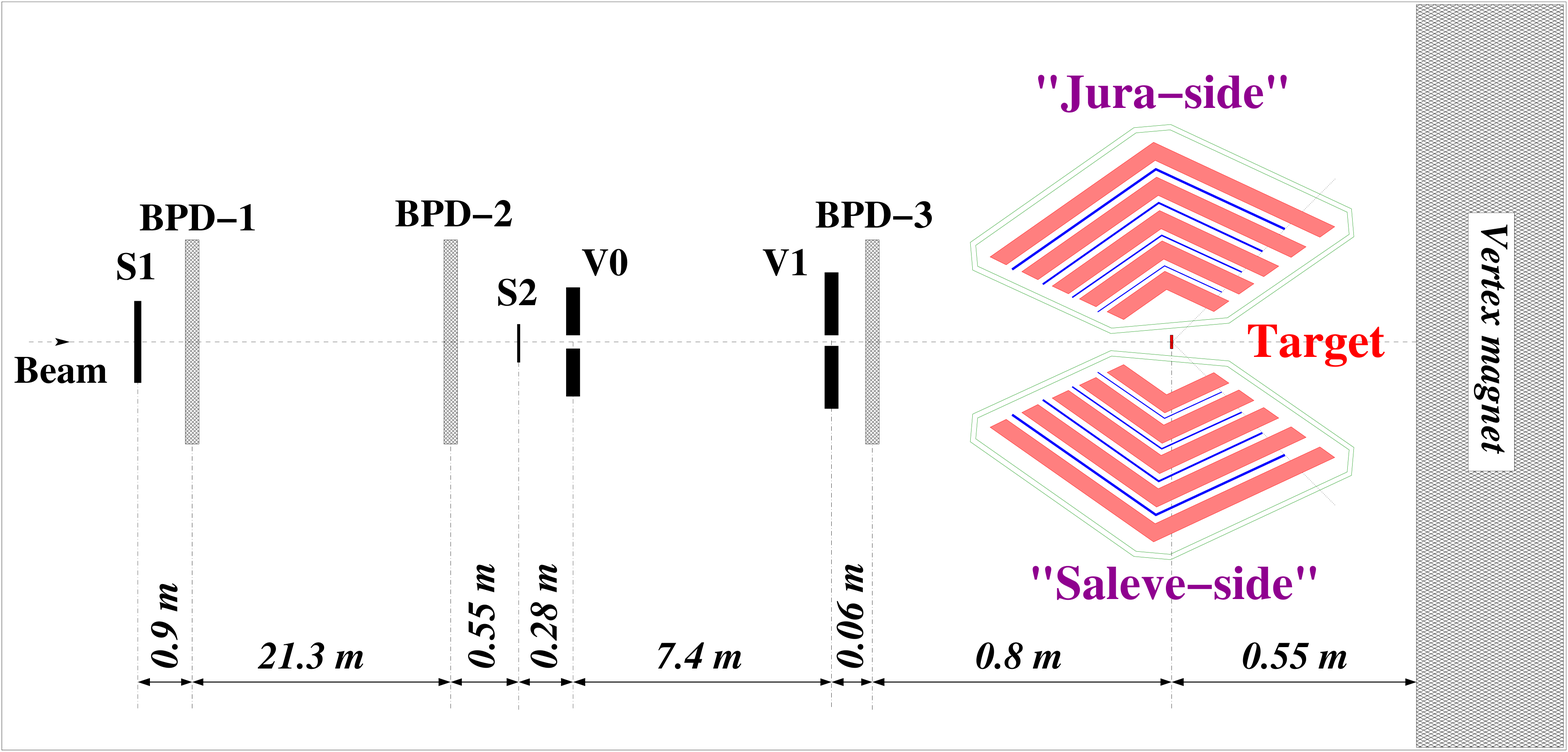}
\caption{Outline of the complete system in downstream (upper panel) and in target position (lower panel; not to scale, note distances indicated).}
\label{fig:LMPD_outline}
\end{center}
\end{figure}

During the 2011 data taking in ``downstream position'' (Figure~\ref{fig:na61setup}), 
the signal from three plastic scintillators were combined, in coincidence with the incoming beam particle 
(defined by the NA61 beam trigger). The last two scintillators were close to the target. 
The one at 40~cm was a 2~cm wide,
5~mm thick disc, whereas the one at 30~cm was 5~cm by 5~cm with 2~mm
thickness. The elimination of beam halo was achieved by an additional
counter (LMPD-V0) in anti-coincidence, right in front of the target. It was 6~cm circular
scintillator, with 1~cm thickness, and with a 5~mm diameter hole in the
middle. The material budget for LMPD-V0 in the hole was minimized to efficiently reduce background. The outline is shown in Figure~\ref{fig:LMPD_outline}. 
The interactions taking place in the target were captured by an
additional 2.5~cm by 3~cm, 2~cm thick scintillator, 4.5 m behind the
target (LMPD-S3). The geometrical alignment of the setup was very critical in reaching
high signal to background ratio, and was carefully verified by beam scans.

In the ``target position'' (Figure~\ref{fig:na61setup}) the trigger definitions of NA61, similar to the 
former NA49 experiment, were used \cite{na49}. The beam was defined by the coincidence of two 
scintillators (S1 and S2 on Figure~\ref{fig:LMPD_outline}) in anticoincidence with two veto counters 
(V0 and V1 in Figure~\ref{fig:LMPD_outline}). To get identified proton beam, a CEDAR Ring Cerenkov 
Counter was used.

\subsection{Target system}
\label{targetsystem}

During the data collecting periods, targets of different atomic number
(A) and thicknesses were used. In order to estimate the background
from non-target interactions, the target was removed regularly
(``target out'' measurements). The switch between target in and out
positions was performed with a remotely controlled pneumatic moving
mechanism, which eliminated the necessity of entering the experimental
area, and hence improving data taking efficiency. During the physics run in 2012 a thin 
Tedlar foil He pipe was installed around the target to reduce the background. 

In the downstream position the beam quality was less
controlled compared to the case of the NA61 nominal target
region. Due to the small diameter of the target, precise alignment of the beam spot, the target 
and the trigger counters (LMPD-V0 and LMPD-S3) was mandatory. Besides optical alignment, 
we have opted for a direct
alignment cross-check based on actual particle data. To this end, the
2010 Prototype was used as a monitor for incoming beam particle
positions for some of the data taking time.

\section{Krypton calibration}
\label{calibration}

For the read-out of our detector 2x9 FE cards are used, 
each of them has 2 amplifier chips with 16 + 16 channels. Since the 
amplification of the chips can be different, the gain may 
vary pad by pad. For the relative gain calibration of the 
pads, random trigger events with $\sp{83}\textrm{Kr}$ source were 
collected. This method came from the ALEPH experiment and 
it was used also in DELPHI and in NA49/NA61 experiments
\cite{na49}.

$\sp{83}\textrm{Kr}$ is an isotope which is produced by electron 
capture from $\sp{83}\textrm{Rb}$. The ground state of $\sp{83}\textrm{Kr}$ is 
not populated directly, the decay chain results a rich structure 
of electron energies in the range of 9--42~keV. During the 
calibration data taking, a foil doped with $\sp{83}\textrm{Rb}$ was 
placed into the existing gas system via a bypass line. 
The gaseous $\sp{83}\textrm{Kr}$ isotope could be easily distributed 
in the chambers, whereas due to the short lifetime of $\sp{83}\textrm{Kr}$, 
no disposal of radioactive gas was necessary and the chambers 
could be operated normally after few half-lives \cite{StarNote}.

\begin{figure}[!ht]
\begin{center}
\includegraphics[angle=0, width=7.5cm]{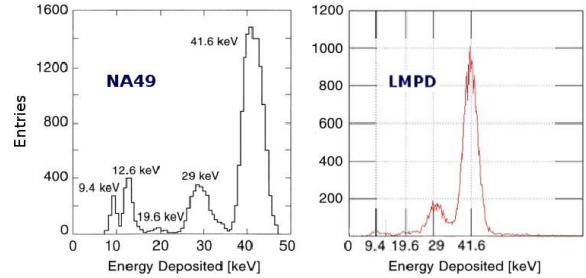}
\caption{Krypton spectra from NA49 (left, \cite{na49}) and measured with LMPD reconstructed by a 3D cluster finder (right).}
\label{fig:Kr-spectra}
\end{center}
\end{figure}

For the analysis of the Kr data collected with LMPD, a ``3D cluster 
finder'' was used which processed in combination the pair of pad rows between two adjacent 
absorbers. This approach was useful in reducing 
charge leakage between the closeby pad rows. The calibration was made 
iteratively, the linearity of the detector response was checked. 
The Kr spectrum from the NA49 experiment and measured with the LMPD 
(on a single pad, reconstructed with the 3D cluster finder) is shown in Figure~\ref{fig:Kr-spectra}.

\begin{figure}[!ht]
\begin{center}
\includegraphics[angle=0, width=7.5cm]{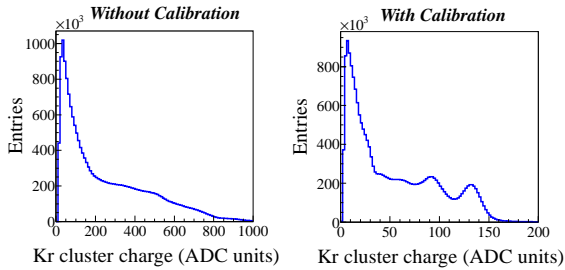}
\caption{Kr spectra on all pads (before and after calibration), reconstructed with the 2D tracking cluster finder.}
\label{fig:spectra}
\end{center}
\end{figure}

\begin{figure}[!ht]
\begin{center}
\includegraphics[angle=0, width=7.5cm]{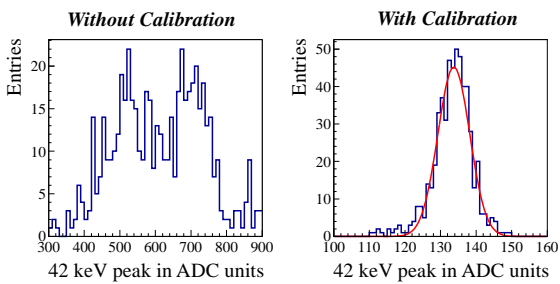}
\caption{Position of the dominant 41.6 keV peak of the Kr spectra in ADC units before and after the calibration. The relative fluctuation is 3.3\% after calibration.}
\label{fig:lastpeak}
\end{center}
\end{figure}

Figure~\ref{fig:spectra} shows the Kr spectrum for all pads in 
LMPD made by the 2D cluster finder (the same as used for the 
analysis of physics data, optimized for tracking) before and after the 
calibration. 
After the calibration the structure of the Kr spectrum is visible, 
the apparent background at low values is a result of charge leakage 
between adjacent pads and pad rows. The position of the 41.6~keV 
peak on each pads is shown on Figure~\ref{fig:lastpeak}. The distribution on the right panel is fitted with a 
Gaussian, resulting in sigma/mean value of 3.3\%. This figure demonstrates the relevance and necessity of 
the Kr calibration, resulting in a highly reliable equalization of the gains.

\section{Performance and pilot data taking results}
\label{results}

\subsection{Event reconstruction and performance}
\label{tracking}

The first step of event reconstruction is the finding of clustered high ADC 
hits on the pad row - time sampling detection planes, which correspond to the 
ionization signals left on a given detection plane (pad row) by the track of the charged 
particles. For this task, we applied a simple closest neighbor search algorithm: if 
any of the charge ADC values on the pad-time plane was at least $C_{\textrm{high}}$=9 
ADC, neighbor search was initiated around it. In case a neighbor had 
at least $C_{\textrm{low}}$=6 ADC charge amplitude, it was considered to belong 
to the same cluster of hits, and its neighbors were also searched for charge above 
the $C_{\textrm{low}}$ threshold recursively. The value of 
$C_{\textrm{low}}$ and $C_{\textrm{high}}$ were a result of optimization, motivated 
by the typical electronic noise level, which was order of $\sigma\approx$ 3 ADC. 
This means that neighbor search was initiated with at least 3$\sigma$ amplitude 
level and was continued recursively with at least 2$\sigma$ amplitude level, whereas 
maxADC of a typical true signal cluster was order of 10$\sigma$ or higher. Therefore, 
these settings are expected to guarantee effective noise rejection along with 
good cluster finding efficiency. 
Indeed, a study of cluster charge distribution with varying 
$C_{\textrm{high}},C_{\textrm{low}}$ values showed that the contribution of 
true signal clusters are not effected by our particular choice of these 
cuts and confirms our expectation. 
After clusterization of hits, the cluster properties are constructed 
by weighted averaging with the charge amplitude. 
In such a way for each cluster the centroid and the elongation parameters on the 
given detection plane are calculated. Elongation parameters are also used for 
electronic noise rejection: due to the construction of the FE electronics 
a typical noise cluster is only one timeslice in time direction but extended 
in the direction of pads, which largely differs from clusters of track signals 
being approximately circular in shape. 
Our particular way of cluster centroid calculation is also commonly 
referred to as center of gravity (COG) method. Studies show \cite{Carbone} 
that centroid estimation algorithms with smaller bias and better resolution 
also exist. However, in our case the simple COG approach was applied 
as the signal clusters consisted of large number of hits with high amplitude, 
the position resolution was dominated by the multiple scattering in the 
absorber layers, furthermore, 
precision tracking was not necessary for our purpose as only multiplicity 
counting was performed in a relatively low population 
detection environment.

\begin{figure}[!ht]
\begin{center}
\includegraphics[angle=0, width=7.5cm]{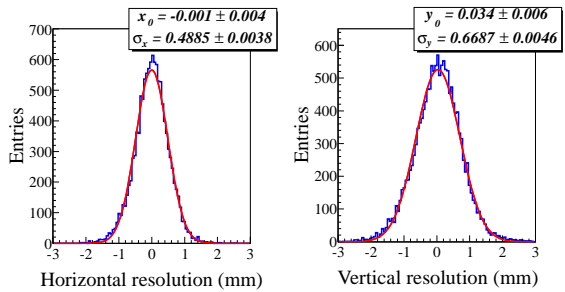}
\caption{Determination of position resolution of cluster centroids via their deviation 
from fitted tracks (histogram: data, solid line: Gaussian fit).}
\label{fig:resolution}
\end{center}
\end{figure}

The second step of the reconstruction is finding of particle trajectories, 
which are straight tracks of clusters in the detection volume. For this task, 
first a simple combinatorial track finder was applied. Clusters starting from 
the target were gathered into track candidates combinatorically, with first 
considering the longer candidates with less number of unregistered intermediate 
clusters. These candidates were fitted with straight line hypothesis assuming 
the same and arbitrary cluster position uncertainty everywhere to construct the 
$\chi^2$ expression to be minimized. The $\chi^2$ distribution of the true 
and false candidates showed a very good separation, and this separation cut 
was used to define accepted candidates. The clusters of the accepted track 
candidates were not considered for the generation of further candidates. 
The distribution of the deviation of 
the cluster centroids from the fitted tracks was used to determine the position 
resolution of the centroid determination method as shown in 
Figure~\ref{fig:resolution}: the position resolution in the pad direction 
was seen to be order of 0.5~mm, while 0.7~mm in the drift direction. 
The measured cluster centroid resolution values 
were used to construct a statistically accurate $\chi^2$ expression for 
track finding and fitting which was then used in the reconstruction of 
the total recorded data.

The combinatorial track finding, however, proved to be very costly 
in computational time in case of events with larger number of clusters, furthermore 
the relative high probability of cluster responses below detection threshold 
posed a complication: one needs to find track patterns with possibly missing 
intermediate measured points while minimizing the inclusion of noise clusters. 
This motivated the development of a track finding method whose cost was not 
increasing factorially with the number of tracks in the event. Our choice 
fell to Hough transformation \cite{Hough} combined with maximum likelihood principle. 

The basic idea of Hough transform is that the position of a cluster centroid 
determines a hyper-plane in the four dimensional vector space of straight 
track parameters through the identities 
$X = M_{X} + Z N_{X}$ and $Y = M_{Y} + Z N_{Y}$, 
where $Z$ is our affine parameter along our track, $M_{X},M_{Y},N_{X},N_{Y}$ 
are our track parameters, while $X,Y$ are the cluster centroid coordinates at a $Z=const$ plane 
through which the track is required to pass. The intersection of such 
hyper-planes determines the straight line tracks. In order to capture the 
described hyper-planes the track parameter space is uniformly binned in each 
direction, the parameters $N_{X},N_{Y}$ are scanned as free parameters of the 
$M_{X} = X - Z N_{X}$, $M_{Y} = Y - Z N_{Y}$ plane, and the corresponding 
intersected $M_{X},M_{Y}$ bins are marked as possible track parameters. 

According to the Hough method, the parameter space bins where lots of hyper-planes 
pass through are considered as track candidates. This method is known to be 
very sensitive to careful choice of parameter space bin size, as with too 
large bins clusters belonging to different tracks may be accidentally merged 
to a single track, while with too small bins only very few planes of the same 
track will intersect in the very same point due to finite resolution of 
cluster centroid positions. Motivated by this, we implemented an improved 
version of Hough transformation.

In the improved version, for each cluster centroid the 
position resolution obtained with the described combinatorial method is also 
used. For each such position measurement $X\pm\sigma{X}$, $Y\pm\sigma{Y}$ 
at $Z=const$ the $\pm3\sigma$ band
$\delta{M}_{X} = 3\sigma{X} + |Z|\delta{N}_{X}$, 
$\delta{M}_{Y} = 3\sigma{Y} + |Z|\delta{N}_{Y}$
around the nominal Hough plane is considered. For each intersection bin of 
these $3\sigma$ plane bands the statistical $\chi^2$ is calculated using the 
error propagation formula
$\sigma^{2}{M}_{X} = \sigma^{2}{X} + \frac{1}{3}|Z|^{2}\delta{N}_{X}^{2}$,
$\sigma^{2}{M}_{Y} = \sigma^{2}{Y} + \frac{1}{3}|Z|^{2}\delta{N}_{Y}^{2}$,
the quantities $\delta{N}_{X}$, $\delta{N}_{Y}$ being the Hough bin size 
along $N_{X}$, $N_{Y}$. The intersection 
bins, i.e. the track candidates, are then ordered according to their number of 
clusters and according to their $\chi^{2}$ likelihood. These candidates are 
accepted with first preferring the longer and bigger likelihood ones, with a 
subsequent removal of their clusters from the Hough table, thus can be regarded 
as a maximum likelihood track finding method. The cost is merely linear in 
number of clusters $\times$ number of Hough plane-band bins of a typical 
cluster. The Hough table is implemented using a container not storing the 
bins unoccupied by clusters, and thus reducing the memory requirement to 
approximately the square-root of the total Hough binning. 

%This principle was realized using the 'map' container of the 
%C++ Standard Library, where the map index was the unique identifier of the 
%occupied Hough space bin. Since the indexing operation of such Standard Library 
%map is logarithmically expensive in terms of the map population, via 
%this solution the necessary memory usage of the Hough table was reduced 
%approximately to its $N_{X},\,N_{Y}$ projection --- at the expense of 
%logarithmically increasing CPU usage when accessing the occupied Hough bins 
%via indexing.

\begin{figure}[!ht]
\begin{center}
\includegraphics[angle=0, width=7.5cm]{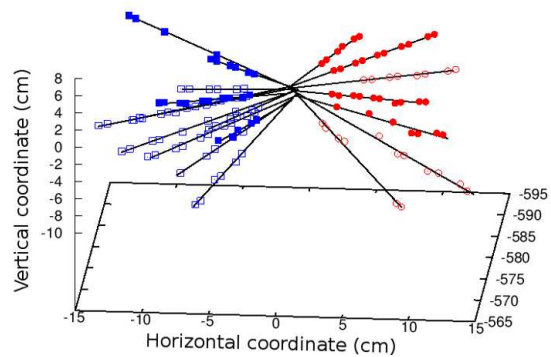}
\caption{Track finding for different number of measured clusters 
(10 superimposed events, points: clusters, lines: tracks).}
\label{fig:event}
\end{center}
\end{figure}

The cluster and reconstruction, 
calibration and analysis software is implemented in the standard offline 
software framework, Shine, of the NA61 experiment \cite{Sipos}.
The performance of the event reconstruction was verified by eye scans over 
sample of 500 events, and proved to be close to ideal. 
Figure~\ref{fig:event} shows track reconstruction in operation for tracks 
with different number of measured clusters.

\begin{figure}[!ht]
\begin{center}
\includegraphics[angle=0, width=6.5cm]{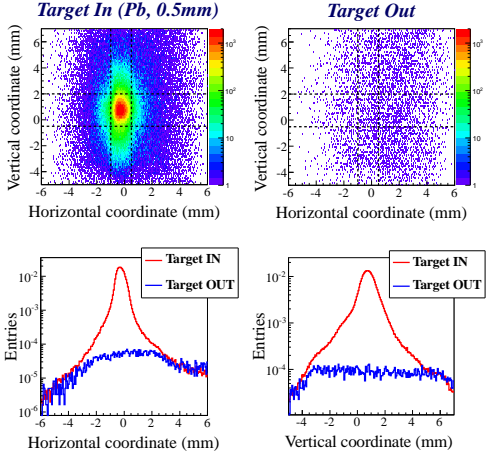}
\caption{Distributions of the extrapolated track points to the target plane (upper left: target in, upper right: target out). Contribution of the target plane is clearly visible. Lower panels show the horizontal and vertical distributions, for the regions indicated by lines on the upper left panel.}
\label{fig:mainvertex}
\end{center}
\end{figure}

After track reconstruction the fitted track may be extrapolated to the constant $Z$ plane 
intersecting with the target. Figure~\ref{fig:mainvertex} shows the 
distributions of these extrapolated intersection point coordinates 
for the target in and the target out data samples. The contribution of interactions 
within the target is clearly visible. The contribution from non-target tracks in the target 
region is well below the percent level, demonstrating the success of background suppression.

\subsection{Ionization for a given range: demonstration of Z=1 particle identification}
\label{identification}

\begin{figure}[!ht]
\begin{center}
\includegraphics[angle=0, width=7.5cm]{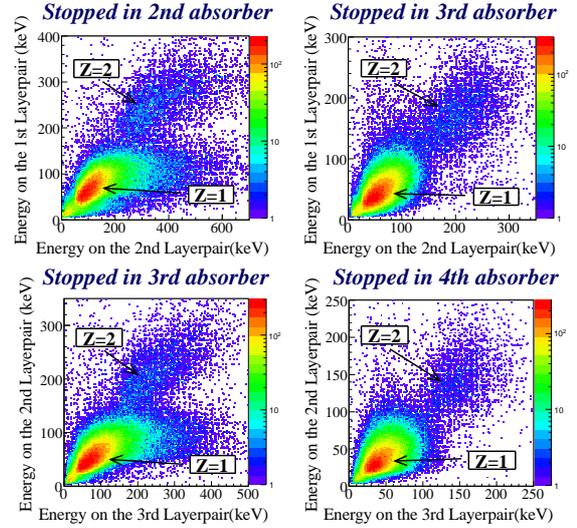}
\caption{Correlation of deposited energy (dE/dx) for stopped particles in adjacent layers. 
Peaks for Z=1 (mainly protons, but also deuterons) and Z=2 (He) are clearly visible.}
\label{fig:charge2d}
\end{center}
\end{figure}

The particle identification concept adapted for the LMPD is the
simultaneous range and ionization (dE/dx) measurement. The former is a
direct result of a reliable tracking algorithm, whereas the latter
requires precise calibration taking into account angular effects as
well. However, already on the level of reconstructed data, the
demonstration of the concept is possible. Figure~\ref{fig:charge2d} top left panel
shows those tracks which have stopped in the second absorber, that is,
measured in the first two detection layers (4 pad rows) without continuation in the sensitive
volume. The ionization added up on the first two pad rows (first
detection layer) correlates well with the ionization on the second
pair of pad rows (second detection layer), and a marked peak around 60
keV matches well with the expected most probable ionization for
protons (see Table~\ref{tab:absorbers}), but it contains also the deuterons. A peak at four times larger ionization corresponds to 
alphas and $\sp{3}\textrm{He}$. 

The other panels of Figure~\ref{fig:charge2d} shows the similar 2 dimensional 
energy deposition distribution for the tracks stopped in the given absorber. 
The Z=1 and Z=2 peaks are visible on all plots.

\subsection{Comparison to PAI simulation of dE/dx}
\label{comparison}

\begin{figure}[!ht]
\begin{center}
\includegraphics[angle=0, width=7.5cm]{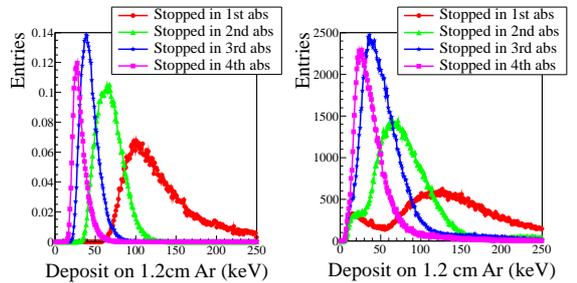}
\caption{Distribution of ionization (dE/dx) for the first measured layer: comparison of
simulation based on the PAI model (left panel) and the measurement (right panel).}
\label{fig:dedx}
\end{center}
\end{figure}

The measured dE/dx distributions in the first detection layer for
angles which are closely perpendicular to the absorbers may be
compared to a simulation based on the PAI model \cite{pai},
calculated by the authors. On the left panel of Figure~\ref{fig:dedx},
the simulation result is shown, for proton tracks which are stopped in any of
the four absorbers. The right panel shows the actual measurement in
the LMPD in a physics run. The differences are due to the fact that the simulation includes only protons, whereas in the measured data protons, deuterons and pions are also pesented. Though this figure serves only for the purpose of a qualitative 
comparison and needs refinements from both the simulation and the data analysis sides,
the similarity is clear, and proves the validity of the proposed PID concept based on
dE/dx and range measurement.

\section{Conclusions}
\label{conclusion}

The paper has presented the design, construction and operation of the
Low Momentum Particle Detector, a new component of the CERN NA61
Experiment. It has been demonstrated that with this small TPC a highly
reliable tracking is possible for tracks emitted from
the target, and high ionization tracks can be tagged as gray particle
candidates. In some momentum ranges, defined by absorbers, particle
identification is directly possible, differentiating Z=1 particles from
pions or heavy fragments. The detector will provide useful input for
understanding slow particle production in hadron-nucleus interactions,
correlating production properties with the production of forward
particles, and especially clarifying the role of ``black'' and ``gray''
protons in collision centrality determination.

\section{Acknowledgements}

The authors acknowledge the financial support from the Hungarian
National Research Fund (OTKA) 68506 and the
Momentum (``Lend\"ulet'') Programme of the Hungarian Academy of Sciences. We wish
to thank for the support of the ``REGaRD'' group of the Wigner RCP in
Budapest, and the technical help from M. Wensween and all members of
the CERN NA61/Shine Collaboration.

\end{document}